\documentclass[12pt]{iopart}

\usepackage{iopams}
\usepackage{setstack}
\usepackage{amscd}
\usepackage{hyperref}
\usepackage{epsf}

\def\be{\begin{equation}}
\def\ee{\end{equation}}
\def\ba{\begin{eqnarray}}
\def\ea{\end{eqnarray}}

\begin{document}

\rightline{\small Journal of Cosmology and Astroparticle Physics 06 (2004) 002  \hfill hep-ph/0312246}

\title{Degeneracy of consistency equations in braneworld inflation}
\author{Gianluca Calcagni}
\address{Dipartimento di Fisica `M. Melloni', Universit\`{a} di Parma, Parco Area delle Scienze 7/A, I-43100 Parma,
 Italy}
\ead{calcagni@fis.unipr.it}

\begin{abstract}
In a Randall--Sundrum type II inflationary scenario we compute perturbation amplitudes and spectral indices up to
 next-to-lowest order in the slow-roll parameters, starting from the well-known lowest-order result for a de
 Sitter brane. Using two different prescriptions for the tensor amplitude, we show that the braneworld consistency
  equations are not degenerate with respect to the standard relations and we explore their observational consequences. It is then shown that, while the degeneracy between high- and low-energy regimes can come from suitable values of the
   cosmological observables, exact functional matching between consistency expressions is plausibly discarded. This result is then extended to the Gauss--Bonnet case.
\end{abstract}


\maketitle


\section{Introduction}

The advent of new Kaluza--Klein models whose features can be tested by modern astrophysical, cosmological and 
ground-based collider observations has triggered interest in those scenarios in which our visible universe
 is embedded in a higher-dimensional spacetime, with extra dimensions that can be `mildly' compact or even non-compact \cite{ADD98}--\cite{RSb}. See, e.g., \cite{rub01,mar03} for a review. In particular, the Randall--Sundrum type II model \cite{RSb} has received much attention since its birth \cite{CGS}--\cite{FTW} because of its rich theoretical structure in a relatively simple conceptual framework. One of the many aspects of the model is the possibility for the cosmological evolution to change substantially from the standard four-dimensional case, since the extra non-compact dimension can communicate via gravitational interaction with the matter confined in the brane universe. Gravity is free to propagate in the bulk, which is assumed to be empty in the simplest scenarios.

According to the present data, the large-scale structure of the universe, as well as the anisotropies of the cosmic microwave background, can be explained by an early stage of accelerated expansion driven by a slow-rolling scalar field, which boosted primordial quantum fluctuations up to cosmological magnitude. The consequent perturbation spectra have been widely studied in the standard 4D scenario thanks to a number of techniques, including the slow-roll (SR) formalism we will adopt here (for a general review and notation, see \cite{lid97}); in particular, it is possible to combine the available observables in relations that are characteristic of the inflationary paradigm and verify them through CMB and sky-survey data.

Braneworld calculations for the perturbation spectra are much more involved because of the complicated geometrical tissue and only quite general formalisms or approximated approaches have been explored so far. In this paper we continue the discussion of the inflationary consistency equations started in \cite{C}, in which next-to-lowest SR order expressions for perturbation amplitudes and indices are constructed from the lowest-order,
de Sitter results known in the literature. Here we will give another prescription for the tensor amplitude motivated by
recent 5D results \cite{KKT}, showing however that there are no new observational features with respect to the
original ansatz. The reliability of the consistency equations for testing high-energy effects is then addressed;
it is shown that experimental degeneracy can come from a low tensor-to-scalar amplitude ratio as well as a nearly scale-invariant spectrum. However, simple slow-roll considerations involving the SR structure of the cosmological large-scale observables show that exact functional degeneracy of at least one relevant consistency equation is expected to be discarded, both in RS and Gauss--Bonnet gravity, whatever the full five-dimensional amplitudes may be.

The plan of the paper is the following: section \ref{setup} sets up the background model; sections \ref{scalar} and \ref{tensor} are devoted to the SR treatment of the cosmological perturbations; in section \ref{conseq} and in the appendix we study the consistency equations and their structure, while the final discussion is in section \ref{disc}.


\section{Basic setup} \label{setup}

The basic model consists in a Randall-Sundrum type II scenario with a brane-confined inflaton field and no energy
 exchange between the brane and the bulk. The continuity equation is the standard 4D equation coming from the
 local conservation of the stress-energy tensor, $\nabla^\nu T_{\mu \nu}=0$:
\be
\dot{\rho}+3H (\rho+p) = 0\,,
\ee
and the modified Friedmann equations in this conformally flat background are \cite{CGS}--\cite{FTW}
\ba
H^2 = \frac{\kappa_4^2}{6 \lambda} \rho (2 \lambda +\rho)\,,\label{FE1}\\
\frac{\ddot{a}}{a} = -\frac{\kappa_4^2}{6 \lambda} [\lambda (\rho+3p)+\rho(2\rho+3p)]\,,\label{FE2}
\ea
where $H$ is the Hubble parameter restricted to the 3-brane, $\kappa_4^2 \equiv 8\pi/m^2_4$ includes the effective
Planck mass $m_4 \simeq 10^{19}\,{\rm GeV}$, $\lambda$ is the brane tension (constrained to be
$\lambda^{1/4}>100\,{\rm GeV}$ by gravity experiments) and $\rho$ is the matter density, which is $\rho = \dot{\phi}^2/2 + V =p+2V$ in the case of a homogeneous scalar field with potential $V(\phi)$. The case of brane-bulk energy flow is far richer and will not be treated here (e.g., \cite{VDMP}--\cite{LMS}). In the following, superscripts (\textit{l}) and (\textit{h}) will denote the low- ($\rho\ll \lambda$) and high- ($\rho \gg \lambda$) energy regimes, respectively.

The slow-roll parameters are
\ba
\epsilon &\equiv& -\frac{\dot{H}}{H^2} \geq 0\,, \label{epsilon}\\
\eta     &\equiv& -\frac{\ddot{\phi}}{H\dot{\phi}} \label{eta}\,,\\
\xi^2    &\equiv& \frac{1}{H^2} \left(\frac{\ddot{\phi}}{\dot{\phi}}\right)^. = \frac{\dddot{\phi}}{H^2\dot{\phi}}-
 \eta^2\,.\label{xi}
\ea
Since $\ddot{a}/a = H^2 (1-\epsilon)$, the condition for inflation is precisely $\epsilon < 1$ and inflation ends
 when $\epsilon=1$. The time derivatives of the SR parameters are
\ba
\dot{\epsilon} &=& 2H\epsilon\left(\frac{1}{q}\,\epsilon-\eta\right)\label{dotepsi}\,,\\
\dot{\eta}     &=& H (\epsilon\eta-\xi^2)\,,\label{doteta}
\ea
where $q^{(l)}=1$ and $q^{(h)}=2$. For future reference we note that, in the high-energy limit,
\be \label{useful}
\dot{\phi}^2 = \frac{2\mu}{\kappa_4^2}H\epsilon=\frac{\rho}{3}~\epsilon\,,
\ee
where $\mu=\sqrt{\lambda\kappa_4^2/6}\,.$


\section{Scalar perturbations} \label{scalar}

The 5D Einstein equations for a brane with an isotropic fluid embedded in an anti-de Sitter bulk are very
complicated due to both the great number of degrees of freedom for the cosmological perturbations and the non-local
physics coming from the possibility for KK gravitational modes to propagate and interact throughout the whole
spacetime. For this reason, while a general setup has been established \cite{KoS1}--\cite{LCML}, a full 5D spectrum
amplitude has not been calculated yet for either scalar or tensor modes, except in the case of some
particular scenarios, for example those in which the brane is de Sitter \cite{KKT,LMW}--\cite{DLMS} or in the large-scale limit \cite{KoS1,lan01,GKLR,WMLL}. In this context the scalar perturbation is found to be the same as the one in the standard four-dimensional background. A way to see this in the slow-roll framework is to note that in the large-wavelength limit, $k\ll aH$, the four-dimensional perturbed metric induced on the brane can be expressed by means of the two longitudinal-gauge-invariant scalar potentials $\Phi_4$ and $\Psi_4$,
\be
\rmd s_4^2\Big|_{\rm brane} \simeq (1+2\Phi_4)~\rmd t^2-a(t)^2(1-2\Psi_4)\delta_{i\!j}~\rmd x^i \rmd x^j\,.
\ee 
In the case of brane--bulk interaction through a non-diagonal stress--energy bulk tensor, the complete continuity equation is $\dot{\rho}+3 H (\rho+p)+ 2 r_B=0$, where $r_B$ is the 05-component of the bulk tensor (see \cite{VDBL}). Then the 5D equations of motion for both the homogeneous background scalar field $\phi_0(t)$ and its fluctuation $\delta\phi(t,\mathbf{ x})$ are
\ba
\fl\ddot{\phi_0}+3H \dot{\phi_0}+ V'(\phi_0)+2 \frac{r_B}{\dot{\phi_0}}=0\,,\label{eom0}\\
\fl\ddot{\delta\phi}-a^{-2}\nabla^2 \delta\phi+3H \dot{\delta\phi}+ V''\delta\phi-(\dot{\Phi_5}+3\dot{\Psi_5})\dot{\phi_0}+2\Phi_5 V' +4 \Phi_5\frac{r_B}{\dot{\phi_0}}=0\,,\label{eom1}
\ea
where $\Phi_5$ and $\Psi_5$ are the gauge-invariant scalar potentials which appear in the full 5D metric and a prime denotes a $\phi$-derivative. By setting an empty bulk ($r_B=0$) one recovers the ordinary 4D equations and in the large-scale limit $\Phi_5 - \Psi_5 \simeq 0 = \Phi_4 - \Psi_4$. At this point one can conjecture that the perturbed effective gravity and scalar actions giving the equations of motion are the usual 4D ones, and performs pure four-dimensional calculations, for example by the methods of \cite{MFB}. Since the explicit form of $H$ is not involved, as one can easily check, one obtains the standard Mukhanov equation \cite{muk85,muk89} 
\be \label{muksc1}
u_{\mathbf k}''+\left(k^2-\frac{z''}{z}\right)u_{\mathbf k}=0\,,
\ee
where primes denote second derivates with respect to conformal time 
\be \label{confor}
\tau=\int \frac{\rmd t}{a} = -\frac{1}{aH(1-\epsilon)}\,,
\ee
and $u_{\mathbf k}$ are the coefficients of the plane wave expansion of the canonical variable $u=-z {\cal R}$. Here, $z = a \dot{\phi}/H$ and ${\cal R}$ is the curvature perturbation on comoving hypersurfaces. Generally, the spectral amplitudes are defined as
\be \label{Agen}
A^2 =\frac{2k^3}{25\pi^2}\frac{|u_{\mathbf k}|^2}{z^2}\,;
\ee
in particular, the scalar amplitude reads \cite{LS}
\[
A_S = \left.\frac{2^{\nu-3/2}}{5\pi} \frac{\Gamma(\nu)}{\Gamma(3/2)}\left(\frac{1}{1-\epsilon}\right)^{1/2-\nu}\frac{H^2}{\dot{\phi}}\right|_{k=aH}\,,
\]
where $\nu=\sqrt{1/4+\tau^2z''/z} \simeq 3/2 +\Or(\epsilon)$ is a combination of SR parameters and the expression is evaluated at the horizon crossing of the perturbation with wavenumber $k$. This amplitude comes from an exact solution of the Mukhanov equation with constant SR parameters. A group of braneworld solutions found in \cite{HaL} will suit well since they approach the power-law inflation \cite{LM} at late times \cite{C} (see also \cite{RL}). Perturbing around one of these solutions and assuming that $\epsilon$, $\eta$ and $\xi$ are small enough (see equations (\ref{dotepsi}) and (\ref{doteta})), it follows that \cite{SL}
\be
S^{1/2} \equiv A_S = \left.[1-(2C+1)\epsilon+C\eta]\frac{1}{5\pi}\frac{H^2}{\dot{\phi}}\right|_{k=aH}\,,\label{A_S}
\ee
where $C\simeq -0.73$ is a numerical constant; by fixing the square bracket equal to 1 one gets the lowest-order expression. As a final remark we note that this is not a true 5D derivation, nor has the full five-dimensional backreaction of the metric been taken into account. So equation (\ref{A_S}) must be regarded as an effective result rather than an exact one. Moreover, for simplicity we have neglected any contribution coming from both the total anisotropic stress and the Weyl tensor (and their fluctuations); however, we would expect this extra-physics not to dramatically modify the above inflationary perturbation spectrum, since it would only affect the small-scale/late-time structures \cite{LCML,GRS,GM,koy03}.

The spectral index is defined as
\be
n_S-1 \equiv \frac{\rmd \ln S}{\rmd \ln k}= \frac{1}{H (1-\epsilon)} \frac{\dot{S}}{S}\,,
\ee
where we have used the exact relation $\rmd \ln k/\rmd\phi=H(1-\epsilon)/\dot{\phi}$. To first SR order, the scalar
index is uniquely defined in both regimes,
\be \label{n_S}
n_S-1 \simeq 2\eta-4\epsilon\,,
\ee
while the energy dependence is absorbed in the second-order part via equation (\ref{dotepsi}) \cite{C}. Defining the running $\alpha_S \equiv
\rmd n_S /\rmd \ln k$, it is easy to see that
\ba
\alpha_S^{(l)} &=& 2(-4 \epsilon^2+5\epsilon\eta-\xi^2)\,,\label{alpl}\\
\alpha_S^{(h)} &=& 2(-2\epsilon^2+ 5\epsilon\eta-\xi^2)\,.\label{alph}
\ea
In order to write these expressions in terms of observables, we will assume that $\xi \ll \epsilon,|\eta|$ or, alternatively, that $\xi$ does not change in the energy scale, $\xi^{(h)} \simeq \xi^{(l)}$.


\section{Tensor perturbations} \label{tensor}

In the case of a de Sitter brane ($\rho+p=0$), it is possible to decouple the zero-graviton mode from the massive
tower of Kaluza--Klein gravitons, thus obtaining a first-order SR expression for the tensor amplitude \cite{LMW}:
\be
T^{1/2} \equiv A_T = \frac{2}{5\sqrt{\pi}} \left.\frac{H}{m_4}F\left(\frac{H}{\mu}\right)\right|_{k=aH}\,, \label{A_T1}
\ee
where \[F(x)=\left[\sqrt{1+x^2}-x^2 \ln \left(1/x+\sqrt{1+1/x^2}
\right)\right]^{-1/2}\,.\] In the limit $\rho \ll \lambda$, $F(x) \simeq 1$ and one recovers the standard spectrum,
 while in the limit $\rho \gg \lambda$,  $F(x) \simeq \sqrt{3x/2}$. One way to state this result is that, to lowest-order SR, the braneworld tensor amplitude is given by the 4D expression under the mapping
\be \label{map1}
h_0:\quad H \mapsto H F(H/\mu)\,. 
\ee
The amplitude (\ref{A_T1}) and its scalar counterpart give rise to a consistency equation which is the same as the standard 4D relation; in order to check whether the degeneracy is an accidental feature of this particular approach, higher-order amplitudes and consistency equations are needed. Since an enhanced calculation in a non-decoupled regime is a very difficult task, truly 5D-computed expressions have been performed only in a variety of limits and approximations, none of which to second order SR.

In \cite{KKT} a perturbed de Sitter brane whose Hubble constant is experiencing a discontinuous variation $\delta H = H_1-H_2$ is studied. A 5D spectral amplitude $T_{5D}$, obtained as an expansion in $\delta H/H \ll 1$, is provided at mildly large scales, $k \ll H^2/\delta H$, and is compared with the 4D braneworld amplitude $T_{eff}$ derived from a generalization of the map (\ref{map1}) acting on the standard amplitude $T_{4D} \propto \tilde{H}^2$, namely,
\be \label{map2}
h_1:\quad \tilde{H} \mapsto \tilde{H} F(\tilde{H}/\mu)\,.
\ee
Here $\tilde{H}=H[1+\Or(\delta H/H)]$ includes the correction to the lowest-order result coming from the 4D zero-mode--zero-mode Bogoliubov coefficients. It turns out that the lowest-order $\Or(1)$ and next-to-lowest order $\Or(\delta H/H)$ effective amplitudes match the five-dimensional result, 
\be \label{eff}
T^{(0)}\equiv T_{eff}^{(0)}=T_{5D}^{(0)}\,,\qquad T_{eff}^{(1)}=T_{5D}^{(1)}\,,
\ee
and $\left|\left(T_{5D}-T_{eff}\right)/T^{(0)}\right| \sim \Or(\delta H/H)^2$ outside the horizon, while inside the horizon there can be significant discrepancies.

\begin{figure}[ht]
\begin{center}
\epsfbox{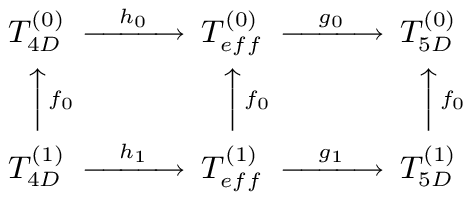}
\end{center}
\caption{\label{fig1} The $h$-mappings (\ref{map1}) and (\ref{map2}) between 4D and effective 5D tensor amplitudes.}
\end{figure}

The case of a continuous smooth variation of the Hubble parameter, which is typical during the inflationary regime long before the reheating, has to be treated separately because of the time dependence of $H$ in the equations of motion of the KK modes. However, one expects the slow-roll generalization of the previous result to display similar features \cite{KKT}, since $\epsilon=\Or(\delta H/H)$. In fact, in order the variation $\delta H$ not to be damped by the accelerated expansion, it must occur in a time interval $\delta t \sim H^{-1}$, that is, $\dot{H} \sim -\delta H / \delta t \sim -\delta H H$, giving the heuristic correspondence $\epsilon \sim \delta H / H$. Thus we can conjecture an analogous mapping (\ref{map2}) in the SR expansion such that (\ref{eff}) holds, with $\tilde{H}=H[1+\Or(\epsilon, \eta)]$. We schematically represent it in figure \ref{fig1}.

Here $f_0$ maps any next-to-lowest-order function to its lowest-order SR form\footnote[1]{This is equivalent to setting $\epsilon=\eta=0$ only in the amplitude $T$, because, for instance, $f_0(S) \propto \epsilon^{-1}$.}. We remark once again that, while $f_0$ is trivial and $g_0 = 1$ was first demonstrated in \cite{LMW}, it would be a non-trivial goal to show that $g_1 = 1$ for a non-de Sitter inflationary brane or, more generally, to find the region in the space of the effective amplitudes where $f_0$ is one-to-one.

There is however another possibility. According to equation (\ref{Agen}), the mapping (\ref{map1}) can be regarded as acting on $z$ rather than on the Hubble parameter:
\be \label{map}
h:\quad z \mapsto z/ F(H/\mu)\,,
\ee
where $z = z_{4D} =\sqrt{2}a/\kappa_4$ for the tensor perturbation. This more elegant formulation has two advantages: first, it encodes all the needed SR information in one single relation, so $h_1$ and $h_2$ collapse to each other; secondly, it is reasonably valid in more general braneworld scenarios \cite{cal5}. 

In \cite{C}, the following ansatz for the tensor amplitude to next-to-lowest order in SR parameters was proposed:
\be \label{Ansatz1}
A_T = \left.\frac{2}{5\sqrt{\pi}}[1-(C+1)\epsilon]\frac{H}{m_4}F\left(\frac{H}{\mu}\right)\right|_{k=aH}\,.
\ee
It is straightforward to see that both the low-energy regime \cite{SL} and the extreme SR regime \cite{LMW} are regained. From the above discussion it emerges that another prescription is viable for the (squared) Randall--Sundrum tensor amplitude: 
\be \label{Ansatz2}
T = T_{4D} \left.\left(\frac{z_{4D}}{z}\right)^2\right|_{k=aH}\,.
\ee
In the high-energy limit,
\be \label{Anshel}
T^{(h)} = \left.\left[1-(3C+2)\epsilon\right]\beta^2 H^3\right|_{k=aH}\,,
\ee
where \[\beta^2 = \frac{1}{25 \pi^2}\frac{3 \kappa_4^2}{4 \mu}\,,\] and we have used the approximated function $F(x) \simeq \sqrt{3x/2}$ together with equations (\ref{dotepsi}), (\ref{confor}) and (\ref{Ansatz2}). The tensor index reads
\be
n_T^{(h)} \equiv \frac{1}{H (1-\epsilon)} \frac{\dot{T^{(h)}}}{T^{(h)}}= -\epsilon[3+(3C+5)\epsilon-2(3C+2) \eta]\,,\label{n_Tnew2}
\ee
to be compared with the low-energy standard result
\be
n_T^{(l)} = -2\epsilon [1+(2 C+3) \epsilon - 2 (C+1)\eta] \label{n_Told2}\,.
\ee


\section{Consistency equations and braneworld degeneracy} \label{conseq}


\subsection{Experimental degeneracy}

In the lowest-order approximation, the only available consistency equation is degenerate, that is,
$n_T^{(h)}=n_T^{(l)}=-2T/S$ \cite{HuL1,HuL2}. Using the relations
\ba
\epsilon \simeq [1-2C(\epsilon-\eta)]\frac{T}{S}\,, &\qquad& {\rm for} \quad\rho \ll \lambda\,,\\
\epsilon \simeq [1-C(\epsilon-2\eta)]\frac{2}{3}\frac{T}{S}\,, &\qquad& {\rm for} \quad\rho \gg \lambda\,,
\ea
we find from equations (\ref{A_S}), (\ref{n_S})--(\ref{alph}), (\ref{Ansatz2}), (\ref{n_Tnew2}) and (\ref{n_Told2}),
\ba
n_T^{(l)} &=& -2 \frac{T}{S}\left[1-\frac{T}{S}+(1-n_S)\right] \label{ecnTold}\,,\\
n_T^{(h)} &=& -2 \frac{T}{S}\left[1-\frac{2}{3}\frac{T}{S}+\frac{2}{3}(1-n_S)\right] \label{ecnTnew}\,,\\
\alpha_S^{(l)} &=& \vphantom{\frac{2}{3}}\,\,\,\,\frac{T}{S} \left[12 \frac{T}{S}+5 (n_S-1)\right]\,, \label{run1}\\
\alpha_S^{(h)} &=& \frac{2}{3} \frac{T}{S} \left[\frac{32}{3} \frac{T}{S}+5 (n_S-1)\right]\,.\label{run2}
\ea
Equation (\ref{ecnTnew}) confirms the idea that the new prescription is more correct than that of \cite{C}, since it bears no trace of the computational method; conversely, in equation (54) of \cite{C} the constant $C$, typical of the solution of the Mukhanov equation, does appear explicitly.

We do not write the consistency equation for the running of the tensor index since it is degenerate and therefore
does not give any additional information. By combining the last four equations, one gets
\ba
\Delta n_T= n_T^{(l)}-n_T^{(h)}              = \frac{2}{3}\frac{T}{S}\left[\frac{T}{S}+(n_S-1)\right]\,,\label{sper1}\\
\Delta\alpha_S=\alpha_S^{(l)}-\alpha_S^{(h)} =
\frac{1}{3}\frac{T}{S}\left[\frac{44}{3}\frac{T}{S}+5(n_S-1)\right]\,.\label{sper2}
\ea 
Note that, provided $T/S+(n_S-1)>0$ and $\alpha_S<0$, the high-energy spectrum has a redder tensor tilt and a steeper scalar running. It was pointed out in \cite{ST} that the failing of the standard lowest-order consistency equation is due to the introduction of the formal limit $\lambda \rightarrow \infty\,,$ which decouples any bulk physics from pure 3-brane quantities but does not necessarily lead to a ``smooth'' connection between 5D gravitational quantities and their 4D equivalent. Indeed the tensor amplitude is a typical example of this mechanism, its new features being encoded both in the explicit $H$-dependence and in the SR structure.

In order to test the observational signatures that the braneworld scenario may produce, we will use the estimates of the
cosmological observables given by the first-year analysis of the WMAP mission \cite{ben03}--\cite{bri03} (see also the
BOOMERanG experiment \cite{deb03} and the Ly-$\alpha$ forest observations \cite{SMM}). Substituting the upper bound for the tensor-to-scalar amplitude ratio $T/S=0.06$ \cite{ben03} and the best fit for the scalar index of \cite{spe03}, $n_S=0.93$, we find from the previous equations 
\be
\Delta\alpha_S \simeq 0.01\,,
\ee
which is close to the error in the estimate of \cite{bri03}, $\alpha_S=-0.04 \pm 0.03$, as well as to the accuracy
 estimate of the future Planck mission \cite{CGL}. The effect coming from the consistency relation for the tensor index is too weak to be detected by the missions of this generation: putting $n_T \simeq -0.1$, equation (\ref{sper1})
 gives $-\Delta n_T/n_T =\Or(10^{-4})$, which does not improve the result coming from equation (\ref{Ansatz1}) \cite{C}. 
 
\begin{figure}[ht]
\begin{center}
\epsfbox{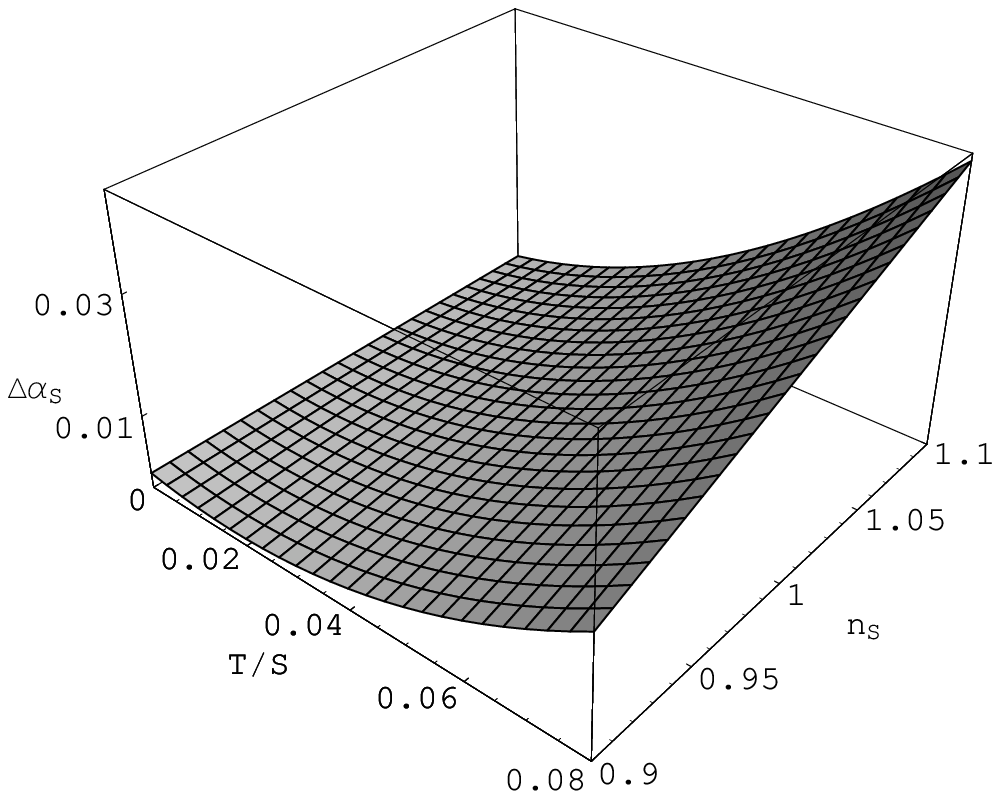}
\end{center}
\caption{\label{fig2} $\Delta\alpha_S$ as a function of the ratio $T/S$ and the scalar index $n_S$.}
\end{figure}

\begin{figure}[ht]
\begin{center}
\epsfbox{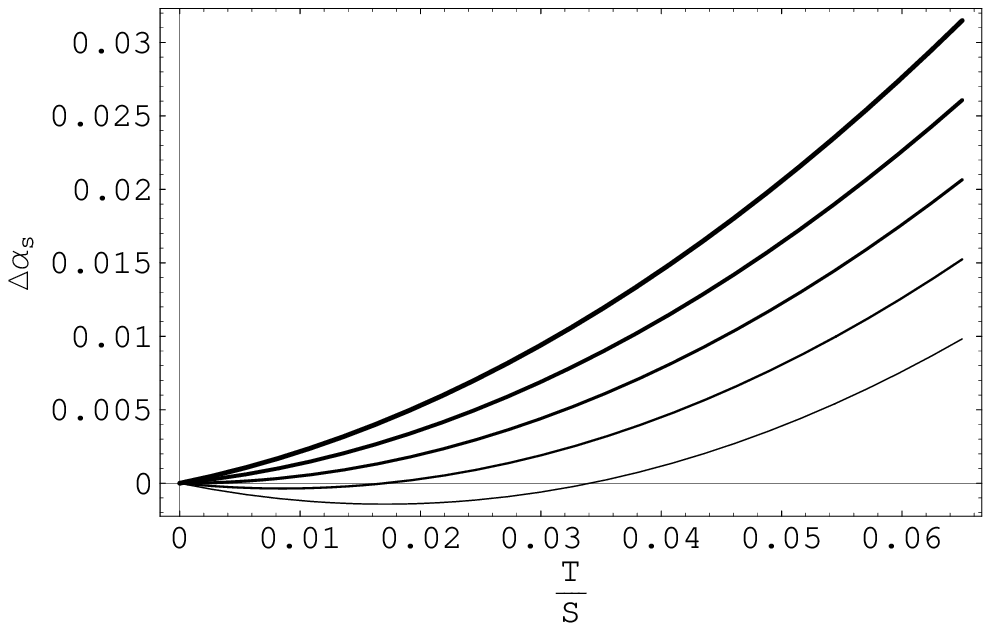}
\end{center}
\caption{\label{fig3} $\Delta\alpha_S$ for particular values of $n_S$. From bottom to top (increasing thickness), $n_S=$ 0.9, 0.95, 1, 1.05 and 1.1.}
\end{figure}

Of course the most practical way to obtain experimental degeneracy is to consider functionally different
consistency equations, whatever they are, in the two regimes and small tensor-to-scalar ratios as well as a
nearly invariant spectrum ($n_S \simeq 1$), an eventuality which is quite possible in the range of the currently available data. For instance, taking the ratio $T/S = (T/S)_{max}/2=0.03$, which is within the $2\sigma$ likelihood bound of \cite{TL}, we have $\Delta\alpha_S \simeq 0.004$ and $-\Delta n_T/n_T = 0.006$, while for $T/S =  (T/S)_{max}/3=0.02$ we get $\Delta\alpha_S \simeq 0.002$ and $-\Delta n_T/n_T = 0.003$, an order of magnitude smaller than the most optimistic high-ratio case. Figure \ref{fig2} shows a three-dimensional plot of equation (\ref{sper2}) as a function of the amplitude ratio and the scalar index, while in figure \ref{fig3} some specific values of $n_S$ are considered. Note that the break of the degeneracy increases for higher values of the spectral index; moreover, the relative sign of the effect may change when $n_S<1$.

\subsection{Theoretical degeneracy}

At this point one may wonder if there can exist next-to-leading-SR-order perturbation amplitudes that give the same
consistency equations for the standard four-dimensional cosmology, thus ruling out the possibility to discriminate between the two scenarios. However, to construct mathematical expressions without any physical content would be of little use since they would not provide a physical explanation why consistency equations should be still degenerate; so the consistency equation approach would still be worth investigation. After this preamble, it
should be noted that it is not possible to construct simple perturbation amplitudes that give degenerate consistency
equations \textit{and} reproduce the extreme SR limit (\ref{A_T1}). The reason is the following. We will concentrate on the consistency equation involving the scalar running $\alpha_S$ since it is the most relevant relation from an observational point of view. It is an equation coming from the lowest-SR-order part of the scalar index, equation (\ref{n_S}), which is generated by the lowest-order part of the scalar amplitude, that is the functional part of (\ref{A_S}). We recall that each time derivative raises the SR order by one at any step: the time derivative of the functional part of the amplitude gives the linear part of the spectral index, while another derivation gives the running, through equations (\ref{dotepsi}) and (\ref{doteta}). Therefore one should change the \emph{functional} part of the amplitude $S$ in order to impose $\alpha_S^{(h)}=\alpha_S^{(l)}$, regardless of the explicit SR structure. Since equation (\ref{A_S}) has not been computed in a truly 5D framework and the bulk source term of (\ref{eom0}) has not been taken into consideration, this could be feasible \emph{a priori}, even if equation (\ref{A_S}) is well supported in many respects\footnote[2]{We stress once again that the lowest-order scalar amplitude can be calculated, in the long wavelength part of the spectrum, by a number of different background-independent techniques.}. An example of a generalized scalar amplitude is (the superscript $(h)$ and evaluation at horizon crossing are understood)
\be
S = [1+f_S(\epsilon, \eta)]\frac{s}{25\pi^2}\frac{H^c}{\dot{\phi}^b}\,,\label{Sgen}
\ee
where $b$ and $c$ are constants, $s$ is a normalization pre-factor and $f_S$ is a linear function of the SR parameters and of any dimensionless combination of cosmological quantities such as $m_4$, $\lambda$, $H$ and its time derivatives.

As has been said in the previous section, the mapping $h$ is well motivated only in the quasi-de Sitter case, so in principle a more general high-energy structure for the tensor amplitude than that of equation (\ref{Ansatz2}) is possible, keeping the functional part constrained by the zero-mode 5D calculation:
\ba \label{Tgen}
T= [1+f_T(\epsilon, \eta)]\beta^2 H^3\,,
\ea
where $f_T$ is the tensor counterpart of $f_S$. For reasons of simplicity, we have dropped the $H$ dependence in the SR functions since it gives rise to a polynomial structure for $S$ which does not change the main argument; see the appendix for details. Also, for the above considerations we can ignore these SR functions.

Equation (\ref{Sgen}) generates the spectral index
\be \label{ngen}
n_S-1 \simeq b \eta-c\epsilon
\ee
and the running
\be \label{alphagen}
\alpha_S \simeq \epsilon [-c\epsilon+(b+2c)\eta]=\epsilon \left[\left(1+\frac{2c}{b}\right)(b\eta-c\epsilon)+\frac{2c^2}{b}\epsilon\right]\,.
\ee
The case $b=0\neq c$ ($c=0\neq b$) is discarded because it is not possible to absorb the $\eta$ ($\epsilon$) dependence of the running into a cosmological observable. The standard case $c=2b=4$ is trivial. Now, assuming 
\be \label{r}
\frac{T}{S} \simeq r \epsilon\,,
\ee
 $r$ being a constant, equation (\ref{run1}) holds in the high-energy limit if $b= 24 r^2/(5r-1)^2$ and $c=12 r^2/(5r-1)\,$. Using equations (\ref{FE1}), (\ref{useful}), (\ref{Sgen}) and (\ref{Tgen}) we have, to lowest SR order,
\be \label{T/Sgen}
\frac{T}{S} \simeq \epsilon \frac{3}{2s} H^{4-c} \dot{\phi}^{b-2} \propto \epsilon^{b/2} \rho^{b/2-c+3}\,;
\ee
this implies that a high-energy relation such that equation (\ref{r}) is possible only in the standard case. In general, a traditional consistency relation will not be obtained in a braneworld context and the brane tension will appear explicitly, as well as the inflaton potential. Suitable powers of these quantities will be set by imposing that the ratio (\ref{T/Sgen}) is dimensionless: if $d_s$ is the dimension of $s$, then it must be $d_s+c-2b=0$. 


\subsection{Generalization to patch cosmology and the Gauss--Bonnet case}

We can generalize the previous arguments to patch cosmology and, in particular, to the Gauss--Bonnet (GB) scenario. In this case the tensor index consistency relation is no longer degenerate (see \cite{DLMS} and \cite{cal3} for details and references). The modified Friedmann equation of the patch approach is $H^2 \propto \rho^q$, $q$ being the same constant as in equation (\ref{dotepsi}), and $\dot{\phi}^2 \propto \rho\epsilon$, as before. For the GB high-energy regime, $q=2/3$. Equation (\ref{alphagen}) becomes
\be \label{alphaq}
\alpha_S \simeq \epsilon \left[\left(1+\frac{2c}{b}\right)(n_S-1)+\left(\theta-1+\frac{2c}{b}\right)c\epsilon\right]\,,
\ee
where $\theta=2(1-q^{-1})=$ $-1$, 1, 0 for $q=2/3$ (GB), 2 (RS) and 1 (4D), respectively; the tensor amplitude is $T \propto H^{2+\theta}$. One may also try to compensate by hand for the $\epsilon$ factor of (\ref{T/Sgen}) with a slightly more general scalar amplitude. Looking at the ratio (\ref{T/Sgen}), it is clear that there could be a scalar amplitude able to absorb the $\epsilon$ and $\rho$ factors, namely,
\be \label{Sgenalt}
S = \frac{s}{25\pi^2}\epsilon^{-\gamma}\frac{H^c}{\dot{\phi}^b}\,,
\ee
where $\gamma$ is a constant. The spectral index and its running are given by equations (\ref{ngen}) and (\ref{alphaq}) when substituting $b$ and $c$ with
\ba
b &\rightarrow& b +2\gamma\,,\nonumber\\
c &\rightarrow& c +\gamma\,.\nonumber
\ea
In particular, equation (\ref{T/Sgen}) now reads
\be
\frac{T}{S} \propto \epsilon^{\gamma+ b/2} \rho^{[2(2q-1)+b-qc]/2}\,;
\ee
in order to satisfy equations (\ref{run1}) and (\ref{r}), we must solve the system
\ba
\gamma+ \frac{b}{2} = 1\,,\nonumber\\
2(2q-1) + b -qc   = 0\,,\nonumber\\
2\frac{c+\gamma}{b+2\gamma}   = 5r-1\,,\nonumber\\
\left(\theta-2+5r \right)(c+\gamma) = 12r^2\,.\nonumber
\ea
For general $q$, there do exist non-trivial solutions. In the Gauss--Bonnet case, however, $r=1$ in equation (\ref{r}), as in four dimensions \cite{DLMS}, and again the system admits no solution; the same considerations apply for the Randall--Sundrum model, $r=3/2$, regardless of the introduction of the new degree of freedom $\gamma$.


\section{Discussion} \label{disc}

A new era of high-precision cosmological observations (WMAP, 2dF \cite{per01}, SDSS \cite{teg03} and others) can now constrain the high-energy physics of braneworld models predicting significant deviations from the standard four-dimensional big bang scenario \cite{LCML,koy03,TL,BFM}--\cite{LT}. In this perspective, it is useful to investigate what are the main features that can discriminate these models from the 4D case; in particular, an early inflationary period experienced by a Randall--Sundrum brane can display different spectra of cosmological perturbations. 

In this paper we have investigated the relations linking the cosmological observables generated by this accelerated expansion. By extending the ``$h$-mapping'' of the de Sitter case to a braneworld scenario with a continuously and smoothly varying Hubble parameter, it is possible to construct a reasonable next-to-lowest-order expression for the tensor amplitude starting from the 4D amplitude of the same order; this procedure leads to a set of consistency equations which carry informations about the extra dimension and that can be probed by the cosmological experiments of this generation. It turns out that, even if the SR structure of the observables is rather rigid, the consistency equations of the standard four-dimensional inflation no longer hold in the brane cosmology and the lowest-order SR degeneracy is softly broken by 5D gravitational effects. 

Therefore, while observational degeneracy of the consistency relations is achievable within the range of cosmological parameters determined by recent experiments, the theoretical structure is unstable against even long-wavelength 5D contributions (via the effective Friedmann equations (\ref{FE1}) and (\ref{FE2})). With explicit examples, we have shown that it would be easy to spoil the simple closed structure of these equations while driving off the standard scalar and tensor high-energy amplitudes, both for Randall--Sundrum and Gauss--Bonnet scenarios. More accurate SR calculations could show consistency relations with a different structure, perhaps unveiling terms that cannot be directly related to the available observables \cite{ST}. Whatever the final answer turns out to be, there is hope of tracking down braneworld signatures through large-scale inflationary physics.


\ack

The author thanks A R Liddle for useful discussions.


\appendix
\section*{Appendix: slow-roll structure of cosmological observables} \label{app1}
\setcounter{section}{1}

In this section we provide a general argument to show that the lowest-order linear structure of the spectral indices and the quadratic one for their running are a consequence of the slow-roll approach, thus justifying the use of a scalar amplitude $S \propto H^c/\dot{\phi}^b$ as a meaningful probe in analysis. This, however, does not say anything new about the consistency relations because their closed structure depends on the form of the tensor-to-scalar ratio, say, equation (\ref{T/Sgen}). First of all, from equations (\ref{epsilon}) and (\ref{dotepsi}) we have
\be
\ddot{H} = H^3 \epsilon (\theta\epsilon+2\eta)\,.\label{ddH}
\ee
Now define a tower of slow-roll (hereafter, `$n$SR') parameters as
\be \label{tow}
\epsilon_n \equiv -\frac{\partial^{n+1} H}{H \partial^{n} H}\,,
\ee
where $\partial^{n} H$ indicates the $n$th time derivative of $H$. In general, this definition would not be convenient for two reasons. The first is that, when expanding around a flat potential, the convergence of the parameters to zero would be made unclear by the derivative in the denominator. The second reason is that it is not possible to truncate any expansion to a finite slow-roll order because all the $n$SR parameters are involved with the same power, unlike what happens with the standard definitions. However, since we are interested in the behaviour of leading-order expressions in the sense of equations (\ref{epsilon})--(\ref{xi}), for our purpose (\ref{tow}) will be sufficient. See \cite{LPB} for a systematic study of the SR tower in the four-dimensional case. Equations (\ref{epsilon}), (\ref{eta}) and (\ref{ddH}) tell us that
\ba
\epsilon &=& \epsilon_0\,,\label{epsi0}\\
2\,\eta &=& -\theta\epsilon_0+\epsilon_1\,,\label{epsi1}
\ea
where $\theta=2(1-q^{-1})$, while, noting that
\ba
\dot{\epsilon}_n &=& H\epsilon_n(\epsilon_0+\epsilon_n-\epsilon_{n+1})\,,\\
\ddot{H} &=& H^3 \epsilon_0 \epsilon_1\,,
\ea
from (\ref{doteta}) we have
\be \label{epsi2}
2\,\xi^2 = \epsilon_1\epsilon_2-\epsilon_1^2+\theta\epsilon_0(\epsilon_0-\epsilon_1)\,.
\ee
In this $n$SR formalism, the explicit energy dependence is removed from the complete slow-roll tower.

In general, a cosmological amplitude $A$ can be written as a sum of dimensionless combinations of $H$ and its derivatives:
\ba
A   &=& \sum_k A_k\,,\\
A_k &=& a_k \prod_{n_k \in U_k} \left(\partial^{n_k}H\right)^{-l_{n_k}}\,,
\ea
where $a_k$ is a normalization pre-factor, $U_k \subset \mathbb{N}$ and the exponents $l_{n_k} \in V_k \subset \mathbb{Q}$ are determined by the physics of the inflationary period. The spectral index is defined as $n_A \equiv (\ln A)^\cdot/H$, that is,
\ba
n_A &=& \frac{\sum_k A_k n_A^{(k)}}{\sum_k A_k}\,,\label{index1}\\
n_A^{(k)} &\equiv& \frac{1}{H} \frac{\dot{A_k}}{A_k}\,.\label{index2}
\ea
Using the distributive property for generic $B_i$s, $(\prod_i B_i)^{\cdot}=\sum_i \dot{B_i} \prod_{j \neq i} B_j\,,$ after some steps one gets
\be
n_A^{(k)} = -\sum_{n_k} l_{n_k} \frac{\partial^{n_k+1} H}{H \partial^{n_k} H} = \sum_{n_k} l_{n_k} \epsilon_{n_k}\,.
\ee
According to this formula, by the very definition (\ref{index1}) the spectral index is linear in the $n$SR parameters $\epsilon_i$ when $l_{n_k}\neq0$ for some $n_k$, since $\Or(n_A)=\Or(n_A^{(k)})\,.$ By equations (\ref{epsi0}) and (\ref{epsi1}), this means that the lowest-order part of the spectral index is at least linear in $\epsilon$ and $\eta$ provided that there are non-zero coefficients\footnote[3]{Equations (\ref{index1}) and (\ref{index2}) are a `first-order' definition with respect to the standard $n_A \simeq (1+\epsilon_0) (\ln A)^\cdot/H$, which has an additional quadratic contribution. So, when we say that the spectral index has an $n$SR linear structure or the running is quadratic, we are dropping the quadratic and cubic standard contribution, respectively.}.

The running of the spectral index is, using (\ref{index1}) and (\ref{index2}),
\be
\alpha_A \equiv \frac{\dot{n}_A}{H} = \frac{\sum_k A_k \alpha_A^{(k)}}{\sum_k A_k}+\frac{\sum_k A_k n_A^{(k)\,2}}{\sum_k A_k}-n_A^2\,,
\ee
where
\be
\alpha_A^{(k)} \equiv \frac{\dot{n}_A^{(k)}}{H}= \sum_{n_k} l_{n_k}\epsilon_{n_k}(\epsilon_0+\epsilon_{n_k}-\epsilon_{n_k+1})\,.
\ee
Again, we find a rigid slow-roll structure that constraints the running to be $\Or(n_A^2)$ or, in the standard SR parameters, to be at least $\Or(n_A^2)$ to leading order.


\end{document}